\documentclass[aps,prb,twocolumn,superscriptaddress,floatfix,longbibliography]{revtex4-2}
\usepackage[T1]{fontenc}
\usepackage[utf8]{inputenc}
\usepackage{graphicx}
\usepackage{hyperref}
\usepackage{amsmath}
\usepackage{amssymb}
\usepackage{amsfonts}
\usepackage{color}
\usepackage{xcolor}
\usepackage{soul}
\usepackage{braket}
\usepackage{upgreek}
\usepackage{babel}
\usepackage[normalem]{ulem}
\hypersetup{linkcolor=red,citecolor=green,urlcolor=magenta,colorlinks=true}

\usepackage{cleveref}
\crefname{appendix}{Appendix}{Appendices}
\crefname{equation}{Eq.}{Eqs.}
\crefname{figure}{Fig.}{Figs.}
\crefname{table}{Table}{Tables}
\crefname{section}{Sec.}{Secs.}

\renewcommand{\d}{\mathrm{d}}
\renewcommand{\epsilon}{\varepsilon}
\renewcommand{\Re}{\mathrm{Re}}

\newcommand{\Tr}{\mathrm{Tr}}
\newcommand{\kk}{\mathbf{k}}
\newcommand{\rr}{\mathbf{r}}
\renewcommand{\AA}{\mathbf{A}} 
\newcommand{\calA}{\mathcal{A}} 
\newcommand{\EE}{\mathbf{E}}
\newcommand{\sgn}{\mathrm{sgn}}

\newcommand{\XX}{\mathbf{X}}
\newcommand{\YY}{\mathbf{Y}}

\makeatletter
\newcommand{\appendixtableofcontents}{\section*{Appendix contents}\@starttoc{apc}}
\newcommand{\addtoappendixtoc}[1]{\addcontentsline{apc}{section}{\protect\numberline{\thesection}#1}}
\makeatother

\begin{document}

\title{Gentle Floquet control of orbital Hall effect and orbital inverse Faraday effect}
\author{Pei-Han Lin}
\affiliation{School of Physics, Peking University, 100871 Beijing, China}
\affiliation{Max-Planck-Institut für Physik komplexer Systeme, 01187 Dresden, Germany}
\author{Roderich Moessner}
\affiliation{Max-Planck-Institut für Physik komplexer Systeme, 01187 Dresden, Germany}
\author{Egor I. Kiselev}
\affiliation{Max-Planck-Institut für Physik komplexer Systeme, 01187 Dresden, Germany}

\begin{abstract}
Floquet engineering controls material responses via a time-periodic electromagnetic drive.
An important task in practice is to reduce the driving strengths required to implement Floquet protocols.
We show that in systems with strong phase-space constraints, where the response is dominated by a small region of momentum space, the relative impact of a resonant Floquet drive is enhanced.
We demonstrate this idea on the example of the orbital Hall effect, where equilibrium occupations restrict the interband coherence to a narrow momentum interval $\hbar\Delta k$. The ratio $eE_0/(\Omega_0\hbar\Delta k)$, where $\Omega_0$ is the frequency of the drive, is considerable even at modest field amplitudes $E_0$, leading to a large relative change in the orbital Hall conductivity.
Due to the non-trivial orbital texture, the circularly polarized drive induces an inverse orbital Faraday effect by magnetizing the electrons. We show that the corresponding susceptibility is controlled by the interband quantum metric that determines the resonant Floquet hybridization.

\end{abstract}

\maketitle

\section{Introduction}

Periodic driving with light is a versatile means for engineering electronic bands and inducing topological phases \citep{oka2009photovoltaic_hall_effect,kitagawa2011floquetinduced,lindner2011floquet,rudner2020band_engineering,kitagawa2010topological_characterization_driven_quantum_system,usaj2014_floquet_graphene_topo,perez2014floquet_traphene_topo,oka2019floquet_review}. Floquet--Bloch states and light-induced Hall responses have been observed experimentally \citep{wang2013_Floquet-Bloch_states_observation,mahmood2016selective_scattering_floquet-bloch_volkov,mciver2020light_anomaouls_Hall_graphene,zhou2023black_phosphorus_floquet,choi2025observation_floquet_graphene,merboldt2025observation_floquet_graphene}. The preparation, occupations, and stability of driven states have motivated extensive work on pump protocols, relaxation, and nonequilibrium steady states \citep{katz2020optically,esin2018q_steady_state_topo_ins,dehghani2015_floquet_topo,genske2015floquet_boltzmann,glazman1983kinetics_pulses_semiconductor,dehghani2014dissipative_topo_floquet,sentef2015pump_probe_floquet,chan2016floquet_ref,farrell2015floquet_ref,gu2011floquet_ref,hubener2017floquet_ref,jiang2011floquet_ref,kennes2019floquet_ref,kundu2013floquet_ref,thakurathi2017floquet_ref,frank2013_lattice_modulations_metal_insulator,castro2022optimal_floquet_control}. Periodic driving has also been used to control correlated and symmetry-broken states \citep{cavalleri2018photo_induced_superconductivity,fausti2011light_induced_superconductivity,forst2011driving_magnetic_order_lattice_excitation,rini2007control_electr_phase_vibrational_excitation,rudner2019_berryogenesis,esin2020floquet_metal_insulator,esin2021,nova2019metastable}.

The field amplitudes required for Floquet engineering depend on the electronic states and observables of interest. 
To modify electronic energies by an amount comparable to a typical bandwidth, the driving field must approach the microscopic electric fields acting on the electrons. Taking an energy of $1\mathrm{eV}$ over a length of $1\mathrm{\mathring A}$ gives the estimate $E_0\sim10^{10}\mathrm{V/m}$ \citep{chaudhary2019orbital}--an exceedingly high value.
Weaker fields can be effective in manipulating low-energy degrees of freedom. Examples are Floquet topological insulators, where circularly polarized light  is used to open topological bandgaps, that can be observed directly \citep{wang2013_Floquet-Bloch_states_observation,zhou2023black_phosphorus_floquet,merboldt2025observation_floquet_graphene,choi2025observation_floquet_graphene} or through transport \citep{mciver2020light_anomaouls_Hall_graphene} at field amplitude of the order of $10^7-10^8\mathrm{V/m}$. Here the Floquet drive competes with broadening due to disorder and finite temperatures. Finally, the impact of Floquet drives can be enhanced by coupling to internal resonances \citep{ron2020ultrafast,disa2021engineering,chaudhary2019orbital,kiselev2024PlasmonMFPD,kiselev2024plasmon_PTC,kiselev2024MagnonMFPD}.

We suggest that systems with strong phase-space constraints, whose transport properties are dominated by small, isolated ``hot regions'' of the Brillouin zone, provide one natural route around this limitation. In such systems, the drive need not substantially modify electronic states throughout the Brillouin zone, but only those within the restricted phase space controlling the response. To explore this idea, we study orbital Hall effects, which are often dominated by relatively small subregions of the Brillouin zone, either due to phase-space constraints or locally pronounced orbital Berry curvature \citep{choi2023OHE_ti}, making them particularly susceptible to phase-space-selective Floquet engineering.

Orbitronics is emerging as a new direction in applied magnetism \citep{fukami2026orbitronics_review,burgos2024_culcer_OAM_review,cysne2025orbitronics_review}. Building on concepts established in spintronics, but relying on orbital degrees of freedom, it promises fast, energy-efficient computation and memory that can be well integrated with existing technology \citep{fukami2026orbitronics_review}. Orbital angular momentum versions of well-known spintronic effects such as the orbital Hall effect \citep{bernevig2005orbitronics,go2018intrinsic} and the orbital Edelstein effect \citep{yoda2018orbital_edelstein,johansson2024theory_edelstein} do not necessarily require strong spin-orbit coupling and thus pose material constraints that are very different from their spin counterparts. They can, for example, be substantial even in light metals \citep{choi2023OHE_ti,lyalin2023observation_Cr,matsumoto2025observation_SI}. The idea of Floquet engineering the orbital Hall effect has been recently proposed \citep{tong2026floquet}; however, we start from very different microscopic assumptions, and show under which circumstances the effects can be particularly large. We focus on the radial--tangential two-band Hamiltonian---a minimal model for momentum-dependent orbital textures and orbital Hall effects \citep{han2023microscopic,tang2024}. Here, Pauli blocking confines the relevant interband processes to a small region in momentum space.

We denote the upper and lower band dispersions by $\epsilon_+(k)$ and $\epsilon_-(k)$ and their Fermi momenta by $k_{F+}$ and $k_{F-}$, respectively. We assume that both bands have positive energies, such that the system has two nested Fermi surfaces (see the inset figure of~\cref{fig:band}). These two concentric Fermi surfaces enclose the ``hot region'' -- an annulus of width
\begin{equation}
\Delta k=k_{F-}-k_{F+},
\end{equation} where the lower band is occupied and the upper band is empty. The intrinsic, equilibrium interband response vanishes outside this annulus, as the bands are either both empty or both full, and the Pauli principle forbids interband transitions.

We consider driving with circularly polarized light of frequency $\Omega_0$ and vector-potential amplitude $A_0=E_0/\Omega_0$. The frequency defines a resonance momentum $k_0$ through
\begin{equation}
\label{resonance_k0_def}
\epsilon_+(k_0)-\epsilon_-(k_0)=\Omega_0.
\end{equation} We will throughout consider the regime $k_{F+}<k_0<k_{F-}$, which places the Floquet avoided crossing inside the annulus with nonzero interband response. It is convenient to define the momentum scale $q_A=eA_0/\hbar$ characterizing the driving amplitude. The central regime of this work is characterized by the two well-separated momentum scales $\Delta k$ and $k_0$, such that
\begin{equation}\label{eq:scale-hierarchy}
q_A\sim\Delta k\ll k_0 .
\end{equation}
The condition $q_A\ll k_0$ ensures that the drive only weakly perturbs the underlying band structure. At the same time, it hybridizes states within a momentum region centered around $k_0$, whose width is of order of $q_A$. Since the interband response of the non-driven system stems from states inside the annulus bounded by $k_{F\pm}$, the relative correction to the orbital Hall conductivity is of order $q_A/\Delta k$. Thus, a drive with $q_A\sim\Delta k$ yields an order-one relative change of the orbital Hall conductivity while remaining weak on the band scale.

\section{Intrinsic orbital Hall effect in the radial--tangential model}
In this section, we establish the orbital response of the undriven system, and identify the phase space that contributes to it. As a prototypical model valid in the vicinity of $\kk=0$, we consider the radial-tangential Hamiltonian \cite{go2018intrinsic,han2023microscopic}. We show that Pauli blocking restricts the interband coherence responsible for the intrinsic orbital Hall effect to an annulus between two Fermi contours, whose width $\Delta k$ is the momentum scale against which the amplitude of the Floquet drive will be compared.

\begin{figure}
\centering
\includegraphics[width=0.45\textwidth]{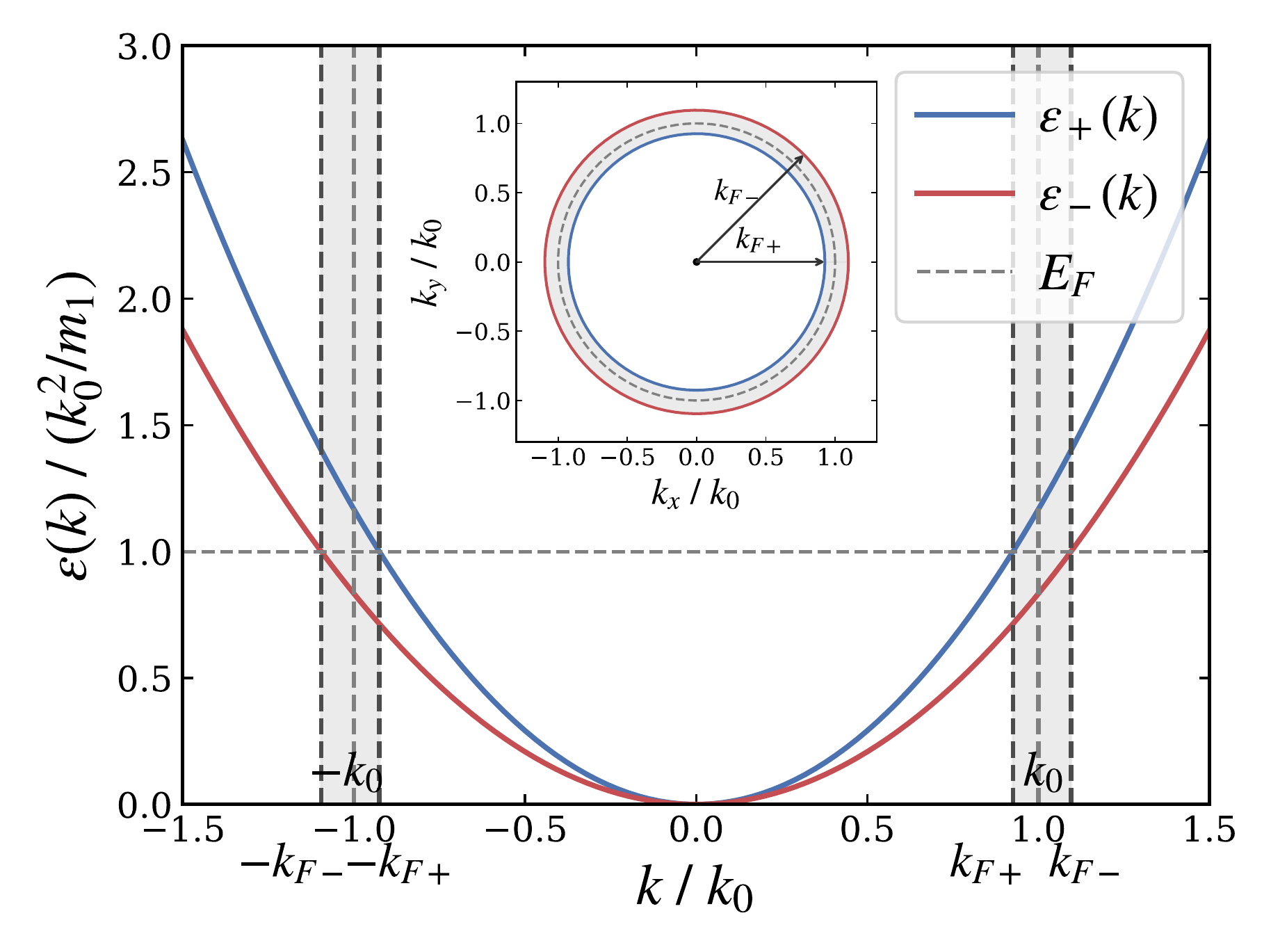}
\caption{Band dispersions $\epsilon_\pm(k)$ of the non-driven Hamiltonian \cref{eq:band_H} for $m_2=6m_1$. The horizontal dashed line marks $E_F$, and the vertical dashed lines mark the Fermi momenta defined in \cref{eq:Fermi-momenta}. The gray intervals, and the gray annulus in the inset, are the hot region $k_{F+}<k<k_{F-}$ in which $f_--f_+=1$, cf.\cref{eq:occupation-window}. The resonance momentum $k_0$ is chosen inside this annulus.}
\label{fig:band}
\end{figure}

The radial--tangential two-band model features two on-site orbitals with identical on-site energies, finite intra- and inter-orbital overlaps, and opposite effective angular momentum quantum numbers. An example would be two dipolar orbitals $p_x$ and $p_y$ with $m_\ell=\pm 1$ on a $C_4$ symmetric lattice. In the Bloch basis
\begin{equation}
|w_{x,y}(\kk)\rangle=\frac{1}{\sqrt{N}}\sum_n e^{i\kk\cdot \mathbf{R}_n}|w_{x,y}(\mathbf{R}_n)\rangle,
\end{equation}
where $w_x$ and $w_y$ are the two orbitals, the Hamiltonian reads \cite{tang2024}:
\begin{equation}\label{eq:band_H}
H_0(\kk)=\begin{bmatrix}
\frac{k_x^2+k_y^2}{m_1}+\frac{k_x^2-k_y^2}{m_2} & \frac{2k_xk_y}{m_2}\\
\frac{2k_xk_y}{m_2} & \frac{k_x^2+k_y^2}{m_1}-\frac{k_x^2-k_y^2}{m_2}
\end{bmatrix},
\end{equation}
where $m_{1,2}$ are the effective masses determined by the orbital overlaps. The eigenvalues (band dispersions) are
\begin{equation}\label{eq:band_dispersion}
\epsilon_\pm(\kk)=\left(\frac{1}{m_1}\pm\frac{1}{m_2}\right)k^2.
\end{equation}
Here $k=|\kk|=\sqrt{k_x^2+k_y^2}$. We use units with $\hbar=1$ from this point onward. Throughout this paper, we focus on the regime $m_2>m_1>0$, where both bands have positive curvature. For a positive Fermi energy $E_F>0$, the model has two concentric, electron-like Fermi contours at
\begin{equation}\label{eq:Fermi-momenta}
k_{F+}=\sqrt{\frac{E_F}{m_1^{-1}+m_2^{-1}}},\qquad
k_{F-}=\sqrt{\frac{E_F}{m_1^{-1}-m_2^{-1}}},
\end{equation}
with $k_{F+}<k_{F-}$. We work at zero temperature with frozen equilibrium occupations. Consequently,
\begin{equation}\label{eq:occupation-window}
f_-(k)-f_+(k)=
\begin{cases}
1,&k_{F+}<k<k_{F-},\\
0,&\text{otherwise}.
\end{cases}
\end{equation}
The band structure and Fermi contours are illustrated in \cref{fig:band}.

The eigenvectors of $H_0(\kk)$, defining the band basis, are
\begin{equation}\label{eq:sbb}
\begin{gathered}
|\psi_+(\kk)\rangle=\cos\theta_k|w_x(\kk)\rangle+\sin\theta_k|w_y(\kk)\rangle,\\
|\psi_-(\kk)\rangle=-i\sin\theta_k|w_x(\kk)\rangle+i\cos\theta_k|w_y(\kk)\rangle,
\end{gathered}
\end{equation}
where $k=|\kk|$ and $\theta_k$ is the polar angle. In this basis, the Hamiltonian reads
\begin{equation}
H_0^\mathrm{band}(\kk)=(m_1^{-1}+m_2^{-1}\tau_3)k^2.
\label{eq:static-band-basis}
\end{equation}

To compute transport, we employ the quantum Boltzmann equation (QBE) for the density matrix $\hat\rho$:
\begin{equation}\label{eq:QBE_general}
\frac{\d\hat\rho}{\d t}=-i[\hat H_0(t),\hat\rho]-i[\hat H_e,\hat\rho]+J[\hat\rho].
\end{equation}
Here $\hat H_e=e\EE_\mathrm{dc}\cdot\rr$ accounts for the external DC electric field $\EE_\mathrm{dc}$. Without loss of generality, we assume that $\EE_\mathrm{dc}=\hat{e}_x E_\mathrm{dc}$. The collision operator $J$ describes scattering processes. 
Since we focus on the intrinsic interband contribution to the orbital Hall response, we evaluate it in the collisionless limit. In the band basis, the collisionless QBE is \cite{tang2024,AHMacdonald2017}:
\begin{equation}\label{eq:QBE_band}
\begin{aligned}
&\frac{\d\rho^\mathrm{band}(\kk,t)}{\d t}=-i[H_0^\mathrm{band}(\kk,t),\rho^\mathrm{band}(\kk,t)]\\
+&e\EE_\mathrm{dc}\cdot\left(\nabla_\kk\rho^\mathrm{band}(\kk,t)-i[\calA(\kk),\rho^\mathrm{band}(\kk,t)]\right),
\end{aligned}
\end{equation}
where the non-Abelian Berry connection is
\begin{equation}
\boldsymbol{\calA}_{ss'}(\kk)=i\langle s\kk|\nabla_\kk|s'\kk\rangle.
\end{equation}
For the band basis in \cref{eq:sbb},
\begin{equation}\label{eq:berryconn}
\boldsymbol{\calA}(\kk)=\frac{\hat{e}_\theta}{k}\tau_1.
\end{equation}
We denote the magnitude of its interband component by $\calA_{+-}(k)\equiv|\boldsymbol{\calA}_{+-}(\kk)|=1/k$.

Solving \cref{eq:QBE_band} with $\d\rho/\d t=0$ yields the steady-state off-diagonal component \cite{tang2024}:
\begin{equation}\label{eq:rhoband-0}
\rho_{+-}^\mathrm{band,(0)}(\kk)=-\frac{\EE_\mathrm{dc}\cdot\boldsymbol{\calA}_{+-}(\kk)}{\epsilon_+(\kk)-\epsilon_-(\kk)}(f_-(\kk)-f_+(\kk)),
\end{equation}
where $f_\pm$ are the equilibrium Fermi functions. By \cref{eq:occupation-window}, $\rho_{+-}^\mathrm{band,(0)}$ is nonzero only in the annulus $k_{F+}<k<k_{F-}$. Thus the annulus of radial width $\Delta k=k_{F-}-k_{F+}$ contains the entire phase space contributing to the intrinsic interband response. This is the origin of the phase-space enhancement described in \cref{eq:scale-hierarchy}: a resonant region of width $q_A$ modifies a fraction of order $q_A/\Delta k$ of the contributing states. In addition, the momentum integral reduces to
\begin{equation}\label{eq:static-response-integral}
\int\frac{\d^2\kk}{(2\pi)^2}(f_--f_+)(\cdots)
=\int_{k_{F+}}^{k_{F-}}\frac{k\d k}{2\pi}
\int_0^{2\pi}\frac{\d\theta_k}{2\pi}(\cdots).
\end{equation}

In the band basis, the orbital magnetization and the longitudinal orbital currents flowing parallel to $\mathbf{E_\mathrm{dc}}$ and orbital Hall currents flowing orthogonal to $\mathbf{E_\mathrm{dc}}$ (see \cref{fig:setup}) are determined by the off-diagonal density matrix element (see Appendix~\ref{sec:observables} for details):
\begin{equation}\label{eq:observables}
\begin{gathered}
m=\mu_B\cdot\int\frac{\d^2\kk}{(2\pi)^2} 2\Re\rho_{+-}^\mathrm{band}(\kk),\\
J_{L_z}^L=\int\frac{\d^2\kk}{(2\pi)^2} \frac{4k_x}{m_1} \Re\rho_{+-}^\mathrm{band}(\kk),\\
J_{L_z}^H=\int\frac{\d^2\kk}{(2\pi)^2} \frac{4k_y}{m_1} \Re\rho_{+-}^\mathrm{band}(\kk).
\end{gathered}
\end{equation}
Substitution of \cref{eq:rhoband-0} into \cref{eq:observables} gives
\begin{equation}
J_{L_z}^{L,(0)}=0,\qquad J_{L_z}^{H,(0)}=\frac{m_2eE_\mathrm{dc}}{2\pi m_1}\ln\frac{k_{F-}}{k_{F+}}.
\end{equation}
The time-reversal-symmetric equilibrium state has zero orbital magnetization and no longitudinal orbital angular momentum response. The intrinsic orbital Hall conductivity is
\begin{equation}
\sigma_{L_z}^{H,(0)}=\frac{J_{L_z}^{H,(0)}}{E_\mathrm{dc}}=\frac{m_2e}{2\pi m_1}\ln\frac{k_{F-}}{k_{F+}}=\frac{m_2e}{4\pi m_1}\ln\frac{m_2+m_1}{m_2-m_1}.
\end{equation}
For $m_1\ll m_2$, $\sigma_{L_z}^{H,(0)}$ approaches $e/(2\pi)$.
For a narrow annulus centered near $k_0$, the logarithm scales as $\ln(k_{F-}/k_{F+})\simeq\Delta k/k_0$. The undriven orbital Hall response is therefore proportional to the small phase-space fraction $\Delta k/k_0$, which makes its relative change particularly sensitive to a drive-induced momentum scale comparable to $\Delta k$.

\section{Resonant Floquet driving of the radial--tangential model}

We now determine how the circularly polarized drive hybridizes the two bands near the one-photon resonance. We first derive the effective Floquet two-band Hamiltonian with its avoided level crossing, and then express the resonant coupling of the two bands in terms of the interband quantum metric. The resulting Floquet states provide the basis for calculating the orbital magnetization and orbital Hall response in the following sections.

\begin{figure}
\centering
\includegraphics[width=0.45\textwidth]{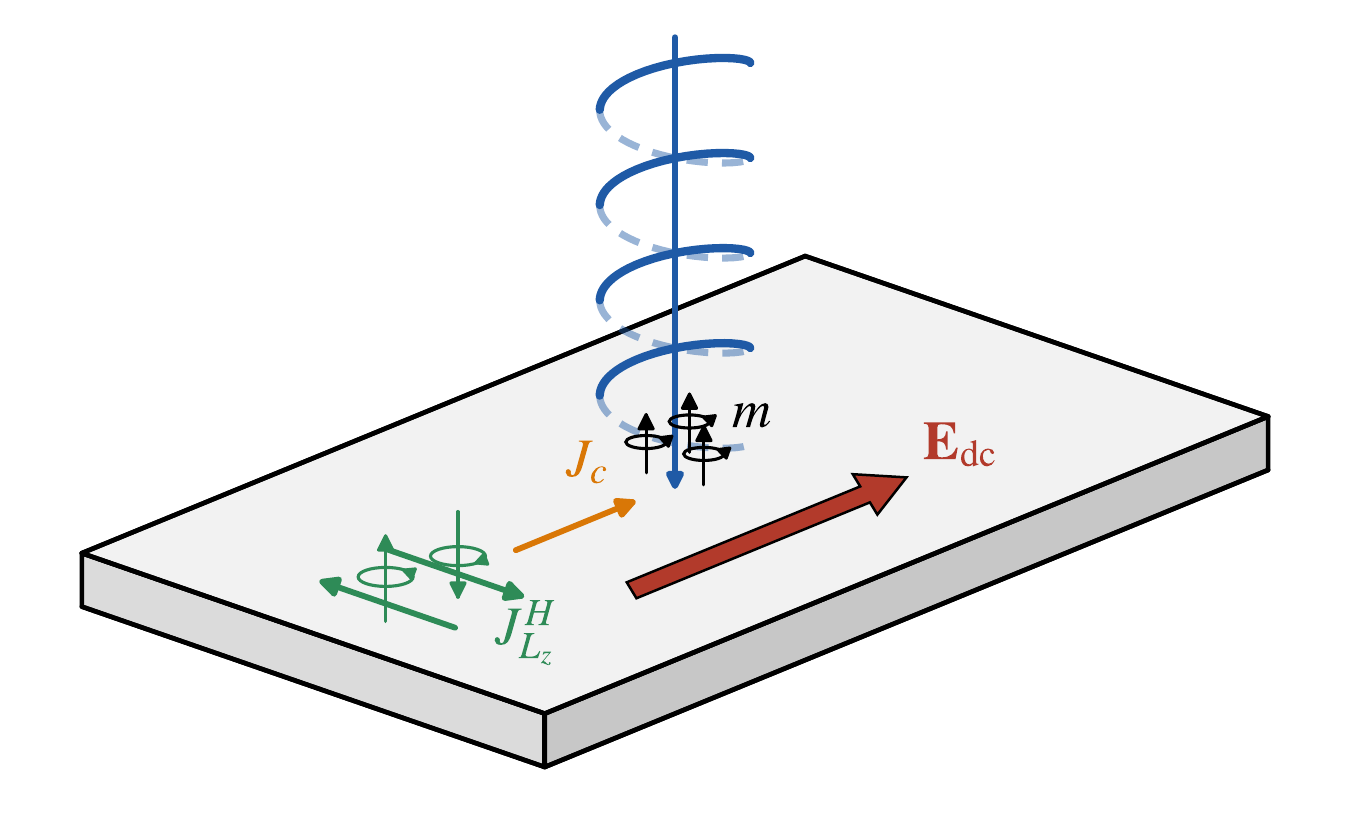}
\caption{The suggested set-up. The blue helix represents the right-handed, circularly polarized AC field, the red arrow represents the DC probe field $\EE_\mathrm{dc}$, and the orange arrow represents the longitudinal charge flow. The circular drive induces a static out-of-plane orbital magnetization $m$ -- a phenomenon known as the inverse Faraday effect \citep{Ziel1965_inverse_faraday_original} -- and modifies the orbital Hall current $J_{L_z}^H$. Reversing the optical helicity reverses $m$.}
\label{fig:setup}
\end{figure}

We introduce an AC electric field characterized by the vector potential $\AA(t)$. The Floquet Hamiltonian is obtained through Peierls substitution. We retain terms linear in $A_0$ in the Hamiltonian. This approximation is controlled by the smallness of $eA_0/k_0$, even when $eA_0$ is comparable to $\Delta k$ as in \cref{eq:scale-hierarchy}. In the band basis, the Hamiltonian reads
\begin{equation}\label{eq:FloquetH_bandbasis}
[H_0^\mathrm{band}(\kk,t)]_{ss'}=\langle \psi_s(\kk)|\hat H_0(\kk+e\AA(t))|\psi_{s'}(\kk)\rangle.
\end{equation}
To linear order in the driving field $\AA$,
\begin{equation}
\left[H_0^\mathrm{band}(\kk,t)\right]_{ss'}=\delta_{ss'}\epsilon_s(\kk)+e\AA(t)\cdot v_{ss'}(\kk).
\end{equation}
Here the velocity matrix element is
\begin{equation}
\begin{aligned}
v_{ss'}(\kk)=&\langle \psi_s(\kk)|\nabla_\kk\hat H_0(\kk)|\psi_{s'}(\kk)\rangle\\
=&\delta_{ss'}\nabla_\kk\epsilon_s(\kk)+i(\epsilon_s(\kk)-\epsilon_{s'}(\kk))\calA_{ss'}(\kk).
\end{aligned}
\end{equation}
As shown in \cref{fig:setup}, we choose right-handed circular polarization, $\AA(t)=A_0(\cos\Omega_0t\hat{e}_x+\sin\Omega_0t\hat{e}_y)$. Without loss of generality, $A_0>0$. The opposite helicity is obtained by changing the sign of the sine term, which reverses the induced orbital magnetization. The Floquet Hamiltonian for the chosen helicity becomes
\begin{equation}
\begin{aligned}
\label{time_dependent_H}
&[H_0^\mathrm{band}(\kk,t)]_{ss'}=\delta_{ss'}\epsilon_s(k)\\
+&eA_0\cos(\Omega_0 t-\theta_k)\delta_{ss'}\frac{\d\epsilon_s(k)}{\d k}\\
+&ieA_0\sin(\Omega_0 t-\theta_k)(\epsilon_s(k)-\epsilon_{s'}(k))\calA_{ss'}(k).
\end{aligned}
\end{equation}

\begin{figure}
\centering
\includegraphics[width=0.45\textwidth]{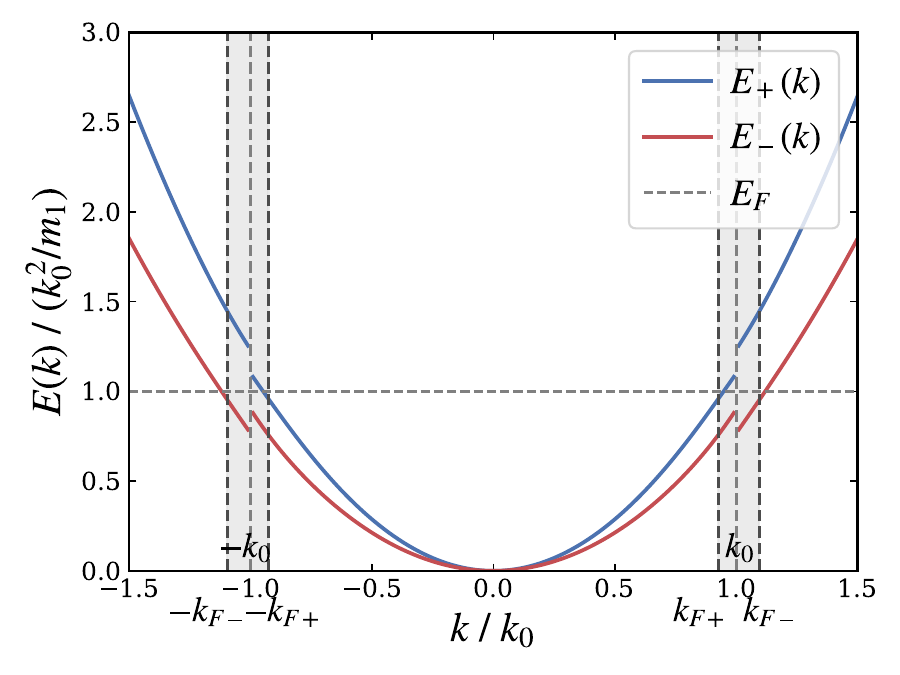}
\caption{Floquet quasienergy bands $E_\pm(k)$ according to \cref{eq:reducible_eigvals}. The vertical dashed lines indicate $k_{F+}$, $k_0$, and $k_{F-}$. The avoided crossing at $k_0$ lies inside the gray annulus $k_{F+}<k<k_{F-}$, where the interband response is nonzero. All parameters are the same as in \cref{fig:band}.}
\label{fig:floquetband}
\end{figure}

The Floquet spectrum is derived in Appendix~\ref{sec:effective}. The one-photon resonance occurs at
\begin{equation}
k_0=\sqrt{m_2\Omega_0/2},
\end{equation}
where $\epsilon_+(k_0)-\epsilon_-(k_0)=2k_0^2/m_2=\Omega_0$. We choose $\Omega_0$ such that $k_{F+}<k_0<k_{F-}$, placing the resonance in the annulus where the interband response is nonzero. The resulting quasienergies are
\begin{equation}\label{eq:reducible_eigvals}
\begin{aligned}
E_+(k)&=\bar\epsilon(k)+\frac{\Omega_0+\Delta(k)}{2},\\
E_-(k)&=\bar\epsilon(k)-\frac{\Omega_0+\Delta(k)}{2}.
\end{aligned}
\end{equation}
Here, $\Delta(k)$ describes avoided crossing induced by the drive:
\begin{equation}\label{eq:signed-floquet-splitting}
\begin{aligned}
\Delta(k)=\sgn(k-k_0)\big[&(\Omega_0-\Delta_0(k))^2\\
&+(eA_0\Delta_0(k)\calA_{+-}(k))^2\big]^{1/2},
\end{aligned}
\end{equation}
Here $\bar\epsilon(k)=\frac{\epsilon_+(k)+\epsilon_-(k)}{2}$ is the band average dispersion, and $\Delta_0(k)=\epsilon_+(k)-\epsilon_-(k)$ is the static band gap. The sign in \cref{eq:signed-floquet-splitting} follows from our choice of Floquet gauge and branch labels. We choose the Floquet representatives on the two sides of the resonance such that $E_\pm$ and $\Psi_\pm$ connect adiabatically to the undriven bands as $A_0\to0$. This requires interchanging the two resonant eigenvectors and shifting one quasienergy by $\Omega_0$ across $k_0$, as shown explicitly in Appendix~\ref{sec:effective}, \cref{eq:Floquet_eigvals_raw,eq:Floquet-branch-choice}. The signed quantity $\Delta(k)$ keeps these band labels continuous through the resonance, while $|\Delta(k)|$ is the magnitude of the avoided-crossing gap. Substituting \cref{eq:band_dispersion,eq:berryconn} gives
\begin{equation}
\Delta(k)=\sgn(k-k_0)\sqrt{\left(\Omega_0-\frac{2k^2}{m_2}\right)^2+\left(\frac{2keA_0}{m_2}\right)^2}.
\end{equation}
The corresponding Floquet dispersion is shown in \cref{fig:floquetband}.

The drive-induced interband coupling can be expressed in terms of the quantum geometric tensor (QGT). For a two-band model, the QGT of band $s$ reads \citep{kolodrubetz2017geometry}
\begin{equation}\label{eq:quantum-geometric-tensor}
Q_{ij}^{(s)}=\langle\partial_{k_i}\psi_s|
(1-|\psi_s\rangle\langle\psi_s|)|\partial_{k_j}\psi_s\rangle
=\calA_{s\bar s,i}\calA_{\bar s s,j}.
\end{equation}
Its real part is the Fubini-Study quantum metric,
\begin{equation}\label{eq:quantum-metric}
g_{ij}^{(s)}=\Re Q_{ij}^{(s)}.
\end{equation}
For a two-band model, the two bands have the same quantum metric, $g_{ij}^{(+)}=g_{ij}^{(-)}=g_{ij}$, while
\begin{equation}
g_{xx}+g_{yy}=\calA_{+-,x}\calA_{-+,x}+\calA_{+-,y}\calA_{-+,y}=\calA_{+-}^2(k).
\end{equation}
Therefore, the Floquet gap can be written as
\begin{equation}
\begin{aligned}
\Delta(k)=\sgn(k-k_0)\big[&(\Omega_0-\Delta_0(k))^2\\
&+(eA_0\Delta_0(k))^2\Tr g(k)\big]^{1/2}.
\end{aligned}
\end{equation}
Here $\Tr g=g_{xx}+g_{yy}$.

\section{Inverse orbital Faraday effect}\label{sec:magnetization}

Circularly polarized light breaks time-reversal symmetry and induces a static orbital magnetization $m$, associated with the inverse orbital Faraday effect \citep{tazuke2025formulation,Sharma2024} (see \cref{fig:setup}). We focus on the atomic contribution to the inverse Faraday magnetization, disregarding itinerant effects \citep{thonhauser2011theory}. The expectation value of $\hat L_z$, time-averaged over one cycle of the drive, is given by
\begin{equation}
\langle\Psi_\pm^\dagger(\kk)\hat L_z\Psi_\pm(\kk)\rangle_\mathrm{cycle}=\pm\frac{eA_0}{2\Omega_0}\frac{\d\Delta_0(k)}{\d k}\sin\phi(k),
\end{equation}
where $\Psi_\pm(\kk)$ are the Floquet eigenstates defined in \cref{eq:Floquet_eigvecs}, and  the mixing angle $\phi(k)$ is defined as
\begin{equation}
\phi(k)=\arctan\frac{eA_0\Delta_0(k)\calA_{+-}(k)}{\Omega_0-\Delta_0(k)},\qquad \phi\in[-\pi/2,\pi/2].
\end{equation}
The derivation is given in Appendix~\ref{sec:OAM}. Substituting \cref{eq:band_dispersion,eq:berryconn} gives
\begin{equation}
\begin{aligned}
&\langle\Psi_\pm^\dagger(\kk)\hat L_z\Psi_\pm(\kk)\rangle_\mathrm{cycle}=\pm\frac{2keA_0}{m_2\Omega_0}\\
\times&\frac{keA_0}{\sqrt{(keA_0)^2+(k_0^2-k^2)^2}}\sgn(k-k_0).
\end{aligned}
\end{equation}
Since the two bands have opposite $L_z$ expectation values, their contributions cancel wherever both are occupied or both are empty. The net magnetization therefore comes only from the annulus defined in \cref{eq:occupation-window} and shown in \cref{fig:band}. For the right-handed polarization used above, the static orbital magnetization is thus
\begin{equation}\label{eq:m-res}
\begin{aligned}
&m=\mu_B\int_{k_{F+}}^{k_{F-}}\frac{k\d k}{2\pi}\int_0^{2\pi}\frac{\d\theta_k}{2\pi}\times \left(-\frac{2keA_0}{m_2\Omega_0}\right)\\
\times&\frac{keA_0}{\sqrt{(keA_0)^2+(k_0^2-k^2)^2}}\sgn(k-k_0),
\end{aligned}
\end{equation}
where $\mu_B$ is the Bohr magneton. The integral in \cref{eq:m-res} is finite throughout the closed interval $k_{F+}\leq k_0\leq k_{F-}$. When the resonance is well separated from both Fermi contours, i.e.
\begin{equation}\label{eq:IFE-asymptotic-condition}
\min(k_0-k_{F+},k_{F-}-k_0)\gg eA_0,
\end{equation}
it has the asymptotic form derived in Appendix~\ref{sec:OAM}:
\begin{equation}\label{eq:m-weak-drive}
m=\frac{\mu_Be^2A_0^2}{2\pi m_2\Omega_0}\left[k_0^2\ln\frac{k_0^2-k_{F+}^2}{k_{F-}^2-k_0^2}-k_{F-}^2+k_{F+}^2\right]+O(A_0^3).
\end{equation} Writing the optical helicity as $\xi=\pm1$, the result for both polarizations is
\begin{equation}
m=\xi\chi_\mathrm{IFE}E_0^2+O(E_0^3),
\end{equation}
where the effective inverse Faraday susceptibility $\chi_\mathrm{IFE}$ is 
\begin{equation}\label{eq:IFE-coefficient-normalized}
\chi_\mathrm{IFE}=\frac{e^2\mu_B}{4\pi\Omega_0^2}\left(\ln\frac{k_0^2-k_{F+}^2}{k_{F-}^2-k_0^2}-\frac{k_{F-}^2-k_{F+}^2}{k_0^2}\right).
\end{equation}
Here we used $\Omega_0=2k_0^2/m_2$ and $E_0=\Omega_0A_0$. The magnetization is quadratic in $E_0$ and reverses sign with the optical helicity. The logarithmic divergence as $k_0$ approaches either Fermi contour marks the breakdown of the asymptotic expansion \cref{eq:IFE-asymptotic-condition}. The exact integral in \cref{eq:m-res} remains finite.

\begin{figure}
\centering
\includegraphics[width=0.45\textwidth]{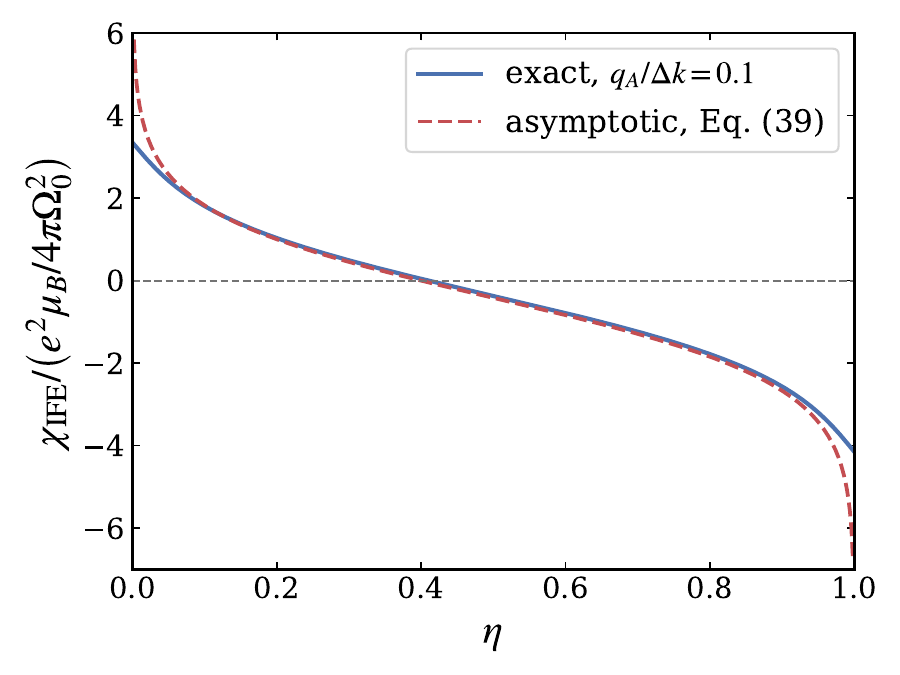}
\caption{Inverse orbital Faraday susceptibility $\chi_\mathrm{IFE}$ in units of $\frac{e^2\mu_B}{4\pi\Omega_0^2}$ as a function of the resonance position $\eta$ defined in \cref{eq:eta-definition}. The solid curve is obtained by integration of \cref{eq:m-res} over the full interval at $q_A/\Delta k=0.1$. The dashed curve is the asymptotic result of \cref{eq:IFE-coefficient-normalized}. We use the same $m_1$, $m_2$, and $E_F$ as in \cref{fig:band}.}
\label{fig:IFE-eta}
\end{figure}

We parameterize the resonance position within the annulus by
\begin{equation}\label{eq:eta-definition}
k_0=\eta k_{F+}+(1-\eta)k_{F-},\qquad 0<\eta<1.
\end{equation}
\cref{fig:IFE-eta} shows $\chi_\mathrm{IFE}$ as a function of $\eta$. It decreases with $\eta$ and changes sign at $\eta\simeq0.410$. The exact finite-drive result approaches the asymptotic curve away from the Fermi contours and remains finite at both endpoints. 

\section{Floquet control of the orbital Hall effect}

We now calculate the orbital Hall conductivity in the presence of the AC drive. To obtain the response linear in the DC probe field $E_\mathrm{dc}$, we substitute the time-dependent Hamiltonian $H_0^\mathrm{band}(\kk,t)$ given in \cref{time_dependent_H} into the QBE \cref{eq:QBE_band}, and retain terms linear in $E_\mathrm{dc}$. We use the frozen equilibrium occupations in \cref{eq:occupation-window}, thereby disregarding heating and the redistribution of occupations between Floquet bands. The full derivation is presented in Appendix~\ref{sec:QBE-solution}. Averaging over one driving period gives the off-diagonal density-matrix element entering the observables of \cref{eq:observables}:
\begin{equation}\label{eq:offdiag}
\begin{aligned}
\Re\langle\rho_{+-}^\mathrm{band}(\kk)\rangle_\mathrm{cycle}=&\frac{keA_0}{m_2\Omega_0}\sin\phi(n_--n_+)\\
-\cos\theta_k eE_\mathrm{dc}\frac{keA_0\sin\phi}{m_2\Omega_0}&\tau_{f,0}\partial_k(n_--n_+)\\
+\sin\theta_k eE_\mathrm{dc}\frac{(n_--n_+)}{k}\bigg(&\frac{\Omega_0\cos\phi-\Delta(1+\cos^2\phi)/2}{\Omega_0^2-\Delta^2}\\
-&\frac{keA_0\sin\phi\cos\phi}{m_2\Omega_0\Delta}\bigg).
\end{aligned}
\end{equation}
Here, $\tau_{f,0}$ is a phenomenological relaxation time, which we introduce to regularize the longitudinal response. We stress that the intrinsic orbital Hall response remains finite in the clean limit $\tau_{f,0}\rightarrow\infty$. The first term in \cref{eq:offdiag} is present at $E_\mathrm{dc}=0$ and is independent of $\theta_k$. It corresponds to the static orbital magnetization discussed in \cref{sec:magnetization}. Setting $n_--n_+=1$ for $k_{F+}<k<k_{F-}$ and substituting into \cref{eq:observables} gives exactly \cref{eq:m-res}. The quantum Boltzmann and Floquet-state calculations of the static magnetization therefore agree.

The second term in \cref{eq:offdiag} is proportional to $\cos\theta_k$ and produces the longitudinal orbital angular momentum current. The third is proportional to $\sin\theta_k$ and produces the orbital Hall current. Because each term is multiplied by $n_--n_+$ or its derivative, its radial integral is confined either to the annulus $k_{F+}<k<k_{F-}$ or to the two Fermi surfaces. Substituting \cref{eq:offdiag} into \cref{eq:observables} gives the conductivities (Appendix~\ref{sec:integral}):
\begin{equation}\label{eq:res}
\begin{gathered}
\sigma_{L_z}^L=\frac{J_{L_z}^L}{E_\mathrm{dc}}=-\frac{e^3\tau_{f,0}A_0^2}{\pi m_1m_2\Omega_0}\left[G_L(k_{F+})+G_L(k_{F-})\right],\\
\sigma_{L_z}^H=\frac{J_{L_z}^H}{E_\mathrm{dc}}=\frac{m_2e}{2\pi m_1}[2G_H(k_0)-G_H(k_{F-})-G_H(k_{F+})].
\end{gathered}
\end{equation}
Here,
\begin{equation}
G_L(k)=\frac{k^4}{\sqrt{(k_0^2-k^2)^2+(eA_0k)^2}},
\end{equation}
and $G_H(k)$ is defined in \cref{eq:GHk-int}. The longitudinal orbital conductivity is proportional to $A_0^2\tau_{f,0}$ and describes the light-induced orbital polarization carried by the longitudinal flow.

The orbital Hall response depends on the position of the interband resonance. For $k_{F+}<k_0<k_{F-}$, the Floquet resonance inside the integration interval produces a correction linear in the drive amplitude $A_0$. If $k_0$ lies outside the annulus $k_{F+}<k<k_{F-}$, the integrand is smooth over the full integration range and the weak-drive correction starts at $A_0^2$. We focus on the first situation. Expanding to leading order gives (see Appendix~\ref{sec:integral})
\begin{equation}\label{eq:OHC-weak-drive}
\sigma_{L_z}^H \approx \frac{m_2e}{2\pi m_1}\left(\ln\frac{k_{F-}}{k_{F+}}-\frac{3eA_0}{2k_0}\right),
\end{equation}
where the linear-in-$A_0$ correction originates from the Floquet resonance inside the annulus of width $\Delta k$. For a narrow annulus centered near $k_0$, $\ln(k_{F-}/k_{F+})\simeq\Delta k/k_0$. Restoring $\hbar$ and recalling that 
\begin{equation}
q_A=eA_0/\hbar,
\end{equation}
\cref{eq:OHC-weak-drive} becomes
\begin{equation}\label{eq:OHC-relative-scaling}
\frac{\sigma_{L_z}^H}{\sigma_{L_z}^{H,(0)}}\simeq1-\frac{3q_A}{2\Delta k}.
\end{equation}
Thus $q_A\sim\Delta k$ produces an order-one relative change while $q_A/k_0\ll1$. The amplitude of an electric field corresponding to $q_A/\Delta k$ is
\begin{equation}\label{eq:electric-field-conversion}
E_0=\Omega_0A_0=\frac{\hbar\Omega_0}{e}\frac{q_A}{\Delta k}\Delta k.
\end{equation}

\begin{figure}
\centering
\includegraphics[width=0.45\textwidth]{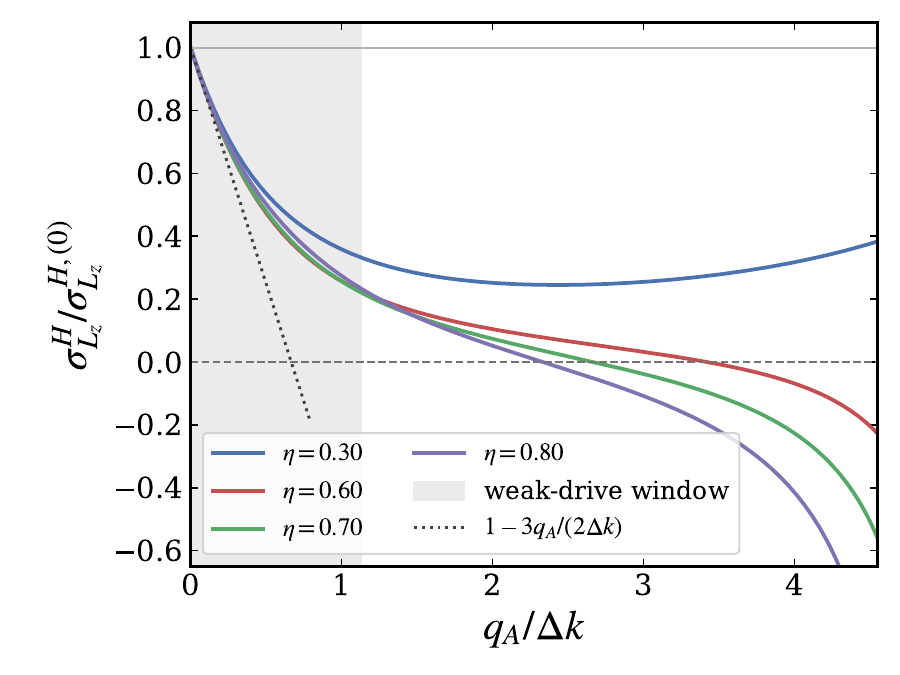}
\caption{Orbital Hall conductivity in the presence of a Floquet drive, cf.~\cref{eq:res}, as a function of $q_A/\Delta k=eA_0/(\hbar\Delta k)$, normalized by its static value. $\eta$, defined in \cref{eq:eta-definition}, describes the position of the resonance. The bandstructure parameters are the same as in \cref{fig:IFE-eta}. For $\hbar\Omega_0=0.2\mathrm{eV}$ and $\Delta k=0.01\mathrm{\mathring{A}}^{-1}$, one unit on the horizontal axis corresponds to $E_0=2.0\times10^7\mathrm{V/m}$ by \cref{eq:electric-field-conversion}. The gray region marks the interval in which $q_A/k_0\leq0.2$ for all plotted values of $k_0$. The dotted line shows the approximated result of \cref{eq:OHC-relative-scaling}. The full result of \cref{eq:res} predicts that the conductivity changes sign depending on the resonance positions at stronger driving.}
\label{fig:OHC-A0dependence}
\end{figure}

The dependence of $\sigma_{L_z}^H$ on $A_0$ is shown in \cref{fig:OHC-A0dependence}. For $m_2=6m_1$, the curves with $\eta=0.60$, $0.70$, and $0.80$ cross zero at $q_A/\Delta k\simeq3.42$, $2.67$, and $2.33$, respectively. The zero crossings lie beyond the weak-drive regime of \cref{eq:OHC-weak-drive}, and the plotted curves use the full expression in \cref{eq:res}. Taking a representative mid-infrared photon energy $\hbar\Omega_0=0.2\mathrm{eV}$ and a narrow-annulus width $\Delta k=0.01\mathrm{\mathring{A}}^{-1}$, \cref{eq:electric-field-conversion} gives $E_0\simeq6.8\times10^7$, $5.3\times10^7$, and $4.7\times10^7\mathrm{V/m}$ for these three zero crossings. These fields are comparable to the $4.0\times10^7\mathrm{V/m}$ peak field used with $191\mathrm{meV}$ photons in Ref.~\cite{mciver2020light_anomaouls_Hall_graphene}. This indicates that while large relative changes in the orbital Hall conductivity can be induced by relatively weak fields, the sign of the Hall conductivity can be reversed at larger, but still realistic field strengths.

\section{Conclusion and Outlook}

Using the radial--tangential model, we have shown that resonant Floquet driving can efficiently modify the orbital Hall conductivity when Pauli blocking confines interband coherence to a narrow region of momentum space. The circularly polarized drive also generates an orbital magnetization whose weak-drive susceptibility is controlled by the quantum geometry of the system.

The present theory could be extended in a number of directions. First, it would be worthwhile to include effects of disorder, finite temperature, heating, and scattering, which will to varying degrees be present in actual experimental realisations.  More broadly, the enhancement of Floquet effects via phase-space constraints can also arise in materials with more complex Fermi surfaces if Pauli blocking confines an interband-coherent response to a narrow momentum region. In more complex settings, additional bands may introduce further resonances and modify the orbital texture and quantum geometry. Furthermore, hot regions with pronounced orbital Berry curvature could lead to similar enhancements. These effects will be considered in future work.

Finally, the strong dependence of the orbital Hall conductivity on the intensity of circularly polarized light is of interest for applications in orbitronics. It may, e.g., enable all-optical orbital-current switches and valves that could control and direct orbital angular momentum currents by locally modifying orbital Hall conductivities with light.

\begin{acknowledgments}
We thank Maximilian Ünzelmann for useful discussions. This project received funding from the Horizon Europe Marie Sk\l odowska-Curie Action program under Grant Agreement 101155351. This work was partially supported by the Deutsche Forschungsgemeinschaft via the cluster of excellence ctd.qmat (EXC 2147, Project-ID 390858490).
\end{acknowledgments}

\bibliography{ref_drivenhall.bib}

\clearpage
\onecolumngrid
\appendix
\appendixtableofcontents

\section{Orbital angular momentum and current operators}\label{sec:observables}
\addtoappendixtoc{Orbital angular momentum and current operators}

The angular momentum operator $\hat L_z$ satisfies
\begin{equation}
\hat L_z|w_x(\kk)\rangle=i|w_y(\kk)\rangle,\qquad \hat L_z|w_y(\kk)\rangle=-i|w_x(\kk)\rangle.
\end{equation}
Therefore,
\begin{equation}
\begin{aligned}
\hat L_z|s=+,\kk\rangle&=i\cos\theta_k|w_y(\kk)\rangle-i\sin\theta_k|w_x(\kk)\rangle=|s=-,\kk\rangle,\\
\hat L_z|s=-,\kk\rangle&=\sin\theta_k|w_y(\kk)\rangle+\cos\theta_k|w_x(\kk)\rangle=|s=+,\kk\rangle.
\end{aligned}
\end{equation}
Its matrix representation in the band basis is therefore
\begin{equation}
L_z^\mathrm{band}=\tau_1.
\end{equation}
Consequently, the orbital magnetization obtained from the density matrix is
\begin{equation}
m=\mu_B\int\frac{\d^2\kk}{(2\pi)^2}\Tr\left[L_z^\mathrm{band}\rho^\mathrm{band}(\kk)\right]=2\mu_B\int\frac{\d^2\kk}{(2\pi)^2}\Re\rho_{+-}^\mathrm{band}(\kk).
\end{equation}

For an internal observable $O$, the corresponding current matrix is
\begin{equation}
\mathbf{J}_O(\kk)=\frac12\{O,\nabla_\kk H_0(\kk)\},
\end{equation}
and its expectation value is
\begin{equation}
\langle\mathbf{J}_O(\kk)\rangle=\Tr[\rho(\kk)\mathbf{J}_O(\kk)].
\end{equation}
Therefore,
\begin{equation}
\begin{aligned}
J_{L_z}^L&=\int\frac{\d^2\kk}{(2\pi)^2}\frac12\Tr\left[\left(L_z\partial_{k_x} H_0(\kk)+\partial_{k_x} H_0(\kk)L_z\right)\rho(\kk)\right],\\
J_{L_z}^H&=\int\frac{\d^2\kk}{(2\pi)^2}\frac12\Tr\left[\left(L_z\partial_{k_y} H_0(\kk)+\partial_{k_y} H_0(\kk)L_z\right)\rho(\kk)\right].
\end{aligned}
\end{equation}
Using $H_0^\mathrm{band}(\kk)=(m_1^{-1}+m_2^{-1}\tau_3)k^2$ gives
\begin{equation}
J_{L_z}^L=\int\frac{\d^2\kk}{(2\pi)^2} \frac{4k_x}{m_1} \Re\rho_{+-}^\mathrm{band}(\kk),\qquad J_{L_z}^H=\int\frac{\d^2\kk}{(2\pi)^2} \frac{4k_y}{m_1} \Re\rho_{+-}^\mathrm{band}(\kk).
\end{equation}

\section{Derivation of the effective Hamiltonian}\label{sec:effective}
\addtoappendixtoc{Derivation of the effective Hamiltonian}

The Floquet Hamiltonian is
\begin{equation}
\left[H_0^\mathrm{band}(\kk,t)\right]_{ss'}=\delta_{ss'}\epsilon_s(k)+eA_0\cos(\Omega_0 t-\theta_k)\cdot\delta_{ss'}\frac{\d\epsilon_s(k)}{\d k}+ieA_0\sin(\Omega_0 t-\theta_k)(\epsilon_s(k)-\epsilon_{s'}(k))\calA_{ss'}(k).
\end{equation}
Its Fourier expansion is
\begin{equation}
H_0^\mathrm{band}(\kk,t)=h_0(\kk)+\sum_{n=1}^\infty \left(h_n(\kk) e^{in\Omega_0 t}+h_n^\dagger(\kk) e^{-in\Omega_0 t}\right),
\end{equation}
with
\begin{equation}
[h_0(\kk)]_{ss'}=\delta_{ss'}\epsilon_s(k),\quad [h_1(\kk)]_{ss'}=\frac12 eA_0 e^{-i\theta_k}\left(\delta_{ss'}\frac{\d\epsilon_s(k)}{\d k}+(\epsilon_s(k)-\epsilon_{s'}(k))\calA_{ss'}(k)\right),\quad h_{n>1}(\kk)=0.
\end{equation}

The solution to the Floquet Schrödinger equation
\begin{equation}
i\frac{\partial\Psi(\kk,t)}{\partial t}=H_0^\mathrm{band}(\kk,t)\Psi(\kk,t)
\end{equation}
is labeled by a Floquet eigenvalue $E$ and a set of Fourier coefficients $\psi_n$:
\begin{equation}\label{eq:Floqet_eigvec_def}
\Psi(\kk,t)=e^{-iE(\kk)t}\sum_{n=-\infty}^\infty \psi_n(\kk) e^{in\Omega_0 t}.
\end{equation}
The Fourier components $\psi_n(\kk)$ satisfy
\begin{equation}\label{eq:Floquet-Schrodinger}
\setlength{\arraycolsep}{1pt}
\begin{bmatrix}
\ddots \\
& h_0+2\Omega_0 & h_1 & 0 & 0 & 0\\
& h_1^\dagger & h_0+\Omega_0 & h_1 & 0 & 0\\
& 0 & h_1^\dagger & h_0 & h_1 & 0\\
& 0 & 0 & h_1^\dagger & h_0-\Omega_0 & h_1\\
& 0 & 0 & 0 & h_1^\dagger & h_0-2\Omega_0 \\
& & & & & &\ddots
\end{bmatrix}
\begin{bmatrix}
\vdots\\ \psi_2\\ \psi_1\\ \psi_0\\ \psi_{-1}\\ \psi_{-2}\\ \vdots
\end{bmatrix}=E\begin{bmatrix}
\vdots\\ \psi_2\\ \psi_1\\ \psi_0\\ \psi_{-1}\\ \psi_{-2}\\ \vdots
\end{bmatrix}.
\end{equation}
Suppressing the momentum arguments and using the real interband connection in \cref{eq:sbb} gives
\begin{equation}\label{eq:FloquetSch}
\begin{bmatrix}
\ddots \\
& \bar\epsilon-\frac{\Delta_0}{2}+2\Omega_0 & -e^{-i\theta_k}V_A & e^{-i\theta_k}V_- & 0 & 0 & 0\\
& -e^{i\theta_k}V_A & \bar\epsilon+\frac{\Delta_0}{2}+\Omega_0 & 0 & e^{-i\theta_k}V_+ & e^{-i\theta_k}V_A & 0\\
& e^{i\theta_k}V_- & 0 & \bar\epsilon-\frac{\Delta_0}{2}+\Omega_0 & -e^{-i\theta_k}V_A & e^{-i\theta_k}V_- & 0\\
& 0 & e^{i\theta_k}V_+ & -e^{i\theta_k}V_A & \bar\epsilon+\frac{\Delta_0}{2} & 0 & e^{-i\theta_k}V_+\\
& 0 & e^{i\theta_k}V_A & e^{i\theta_k}V_- & 0 & \bar\epsilon-\frac{\Delta_0}{2} & -e^{-i\theta_k}V_A\\
& 0 & 0 & 0 & e^{i\theta_k}V_+ & -e^{i\theta_k}V_A & \bar\epsilon+\frac{\Delta_0}{2}-\Omega_0\\
& & & & & & & \ddots
\end{bmatrix}
\begin{bmatrix}
\vdots\\ \psi_{2,-}\\ \psi_{1,+}\\ \psi_{1,-}\\ \psi_{0,+}\\ \psi_{0,-}\\ \psi_{-1,+}\\ \vdots
\end{bmatrix}=E
\begin{bmatrix}
\vdots\\ \psi_{2,-}\\ \psi_{1,+}\\ \psi_{1,-}\\ \psi_{0,+}\\ \psi_{0,-}\\ \psi_{-1,+}\\ \vdots
\end{bmatrix}.
\end{equation}
Here $\bar\epsilon(k)=[\epsilon_+(k)+\epsilon_-(k)]/2$, $\Delta_0(k)=\epsilon_+(k)-\epsilon_-(k)$, $V_\pm=(eA_0/2)\d\epsilon_\pm(k)/\d k$, and $V_A=eA_0\Delta_0(k)\calA_{+-}(k)/2$.

Near resonance, where $\Delta_0\approx\Omega_0$, the central $2\times2$ block of \cref{eq:FloquetSch} is
\begin{equation}
\begin{bmatrix}
\bar\epsilon-\frac{\Delta_0}{2}+\Omega_0 & -e^{-i\theta_k}V_A\\
-e^{i\theta_k}V_A & \bar\epsilon+\frac{\Delta_0}{2}
\end{bmatrix}\begin{bmatrix}
\psi_{1,-}\\\psi_{0,+}
\end{bmatrix}=E\begin{bmatrix}
\psi_{1,-}\\\psi_{0,+}
\end{bmatrix}.
\end{equation}
The solutions are
\begin{equation}\label{eq:order0-eigs}
\begin{gathered}
E_+'=\bar\epsilon+\frac{\Omega_0}{2}+\sqrt{\frac{(\Omega_0-\Delta_0)^2}{4}+V_A^2},\qquad \begin{bmatrix}\psi_{1,-}\\\psi_{0,+}\end{bmatrix}_{+}=\begin{bmatrix}\cos(\phi'/2)\\-\sin(\phi'/2)e^{i\theta_k}\end{bmatrix},\\
E_-'=\bar\epsilon+\frac{\Omega_0}{2}-\sqrt{\frac{(\Omega_0-\Delta_0)^2}{4}+V_A^2},\qquad \begin{bmatrix}\psi_{1,-}\\\psi_{0,+}\end{bmatrix}_{-}=\begin{bmatrix}\sin(\phi'/2)\\\cos(\phi'/2)e^{i\theta_k}\end{bmatrix}.
\end{gathered}
\end{equation}
Here the mixing angle is
\begin{equation}\label{eq:angle-raw}
\phi'=\arctan\frac{2V_A}{\Omega_0-\Delta_0},\quad \phi'\in[0,\pi].
\end{equation}

The central subspace $\{|-,1\rangle,|+,0\rangle\}$ couples through the intraband matrix elements to $|-,2\rangle$ and $|+,-1\rangle$. These two couplings are represented by
\begin{equation}
\begin{bmatrix}
e^{-i\theta_k}V_- & 0\\
0 & e^{-i\theta_k}V_+
\end{bmatrix}.
\end{equation}
The interband matrix element also couples the central subspace to $|+,2\rangle$ and $|-,-1\rangle$, represented by
\begin{equation}
\begin{bmatrix}
0 & 0\\
e^{-i\theta_k}V_A & 0
\end{bmatrix}.
\end{equation}
These four states are separated from the central subspace by energies of order $\Omega_0$ or $2\Omega_0$. We therefore evaluate the energy denominators at $A_0=0$. Their drive-induced corrections enter at higher order. To first order in $A_0$, perturbation theory gives
\begin{equation}\label{eq:amps-perturb}
\begin{gathered}
\psi_{2,+}=-\frac{e^{-i\theta_k}V_A}{2\Omega_0}\psi_{1,-},\quad \psi_{2,-}=-\frac{e^{-i\theta_k}V_-}{\Omega_0}\psi_{1,-},\quad \psi_{1,+}=-\frac{e^{-i\theta_k}V_+}{\Omega_0}\psi_{0,+},\\
\psi_{0,-}=\frac{e^{i\theta_k}V_-}{\Omega_0}\psi_{1,-},\quad\psi_{-1,+}=\frac{e^{i\theta_k}V_+}{\Omega_0}\psi_{0,+},\quad\psi_{-1,-}=\frac{e^{i\theta_k}V_A}{2\Omega_0}\psi_{0,+}.
\end{gathered}
\end{equation}

Using the definition \cref{eq:Floqet_eigvec_def} of the Floquet eigenstate, we combine the amplitudes in \cref{eq:order0-eigs,eq:amps-perturb} and obtain 
\begin{equation}\label{eq:Floquet_eigvecs_raw}
\Psi_+'=e^{-iE_+'t}\begin{bmatrix}
-\sin(\phi'/2)e^{i\theta_k}(1+iK_++L_2'^*)\\
\cos(\phi'/2)e^{i\Omega_0t}(1+iK_--L_1'^*)
\end{bmatrix},\qquad\Psi_-'=e^{-iE_-'t}\begin{bmatrix}
\cos(\phi'/2)e^{i\theta_k}(1+iK_+-L_1')\\ \sin(\phi'/2)e^{i\Omega_0t}(1+iK_-+L_2')
\end{bmatrix}.
\end{equation}
Here
\begin{equation}\label{eq:KL-def}
K_\pm=\frac{2V_\pm}{\Omega_0}\sin(\theta_k-\Omega_0t),\qquad L_1'=\frac{V_A}{2\Omega_0}\tan\frac{\phi'}{2}e^{-2i(\theta_k-\Omega_0t)},\qquad L_2'=\frac{V_A}{2\Omega_0}\cot\frac{\phi'}{2}e^{2i(\theta_k-\Omega_0t)}.
\end{equation}

As $A_0\to0$, the eigenvalues $E_\pm'$ defined in \cref{eq:order0-eigs} approach Floquet-zone-shifted branches. Floquet periodicity identifies $E$ and $E+n\Omega_0$ ($n\in\mathbb{Z}$) as equivalent eigenvalues, with eigenstates $\psi$ and $e^{-in\Omega_0t}\psi$, respectively. Define $k_0$ as the resonance momentum where $\Omega_0=\Delta_0(k_0)$. We fix the Floquet gauge and branch labels by requiring $E_\pm$ and $\Psi_\pm$ to connect adiabatically to the undriven upper and lower bands, respectively, as $A_0\to0$. Because the two resonant branches exchange their undriven-band character at $k_0$, this convention selects different representatives of the raw Floquet solutions for $k<k_0$ and $k>k_0$. The resulting eigenvalues are
\begin{equation}\label{eq:Floquet_eigvals_raw}
E_+=\begin{cases}
\bar\epsilon+\frac{\Omega_0}{2}-\sqrt{\frac{(\Omega_0-\Delta_0)^2}{4}+V_A^2},\qquad k<k_0;\\
\bar\epsilon+\frac{\Omega_0}{2}+\sqrt{\frac{(\Omega_0-\Delta_0)^2}{4}+V_A^2},\qquad k>k_0.
\end{cases}\qquad
E_-=\begin{cases} \bar\epsilon-\frac{\Omega_0}{2}+\sqrt{\frac{(\Omega_0-\Delta_0)^2}{4}+V_A^2},\qquad k<k_0;\\ \bar\epsilon-\frac{\Omega_0}{2}-\sqrt{\frac{(\Omega_0-\Delta_0)^2}{4}+V_A^2},\qquad k>k_0.
\end{cases}
\end{equation}
The corresponding Floquet-gauge representatives are
\begin{equation}\label{eq:Floquet-branch-choice}
\Psi_+=\begin{cases}
\Psi_-',\qquad k<k_0;\\
\Psi_+',\qquad k>k_0.
\end{cases}\qquad \Psi_-=\begin{cases}
\Psi_+'e^{-i\Omega_0t},\qquad k<k_0;\\ \Psi_-'e^{-i\Omega_0t},\qquad k>k_0.
\end{cases}
\end{equation}

The sign in the definition
\begin{equation}
\Delta(k)=\sgn(k-k_0)\cdot\sqrt{(\Omega_0-\Delta_0)^2+4V_A^2}=\sgn(k-k_0)\sqrt{(\Omega_0-\Delta_0)^2+(eA_0\Delta_0(k)\calA_{+-}(k))^2},
\end{equation}
therefore records the exchange of Floquet representatives across the resonance. The branch convention in \cref{eq:Floquet-branch-choice} makes $\Delta(k)$ a signed quantity. The physical gap is $|\Delta(k)|$. This definition combines the $k<k_0$ and $k>k_0$ branches of \cref{eq:Floquet_eigvals_raw} into
\begin{equation}\label{eq:Floquet_eigvals}
E_\pm(k)=\bar\epsilon(k)\pm\frac{\Omega_0+\Delta(k)}{2}.
\end{equation}
Similarly, the redefinition
\begin{equation}
\phi=\arctan\frac{2V_A}{\Omega_0-\Delta_0},\quad \phi\in[-\frac{\pi}{2},\frac{\pi}{2}]
\end{equation}
together with
\begin{equation}
L_1=\frac{V_A}{2\Omega_0}\tan\frac{\phi}{2}e^{-2i(\theta_k-\Omega_0t)},\qquad L_2=\frac{V_A}{2\Omega_0}\cot\frac{\phi}{2}e^{2i(\theta_k-\Omega_0t)}
\end{equation}
yields
\begin{equation}\label{eq:Floquet_eigvecs}
\Psi_+=e^{-iE_+t}\begin{bmatrix}
-\sin(\phi/2)e^{i\theta_k}(1+iK_++L_2^*)\\
\cos(\phi/2)e^{i\Omega_0t}(1+iK_--L_1^*)
\end{bmatrix},\qquad\Psi_-=e^{-iE_-t}\begin{bmatrix}
\cos(\phi/2)e^{i\theta_k}(1+iK_+-L_1)\\ \sin(\phi/2)e^{i\Omega_0t}(1+iK_-+L_2)
\end{bmatrix}.
\end{equation}

\section{Cycle-averaged orbital angular momentum and static magnetization}\label{sec:OAM}
\addtoappendixtoc{Cycle-averaged orbital angular momentum and static magnetization}

The expectation values of $\hat L_z$ in the Floquet states of \cref{eq:Floquet_eigvecs} are
\begin{equation}
\begin{aligned}
&\Psi_\pm^\dagger\hat L_z\Psi_\pm\\
=&\mp\sin\frac{\phi}{2}\cos\frac{\phi}{2}\left(e^{i(\theta_k-\Omega_0t)}(1+iK_++L_2^*)(1-iK_--L_1)+e^{-i(\theta_k-\Omega_0t)}(1-iK_+-L_1^*)(1+iK_-+L_2)\right)\\
=&\mp\sin\frac{\phi}{2}\cos\frac{\phi}{2}\left(e^{i(\theta_k-\Omega_0t)}(1+iK_+-iK_--L_1+L_2^*)+e^{-i(\theta_k-\Omega_0t)}(1-iK_++iK_--L_1^*+L_2)\right)+O(A_0^2)\\
=&\pm\left((K_+-K_-)\sin\phi\sin(\theta_k-\Omega_0t)-\sin\phi\cos(\theta_k-\Omega_0t)+\sin\phi\Re((L_1-L_2^*)e^{i(\theta_k-\Omega_0t)})\right)+O(A_0^2).
\end{aligned}
\end{equation}
Using \cref{eq:KL-def},
\begin{equation}
\Psi_\pm^\dagger\hat L_z\Psi_\pm=\pm\bigg(2\sin\phi\frac{V_+-V_-}{\Omega_0}\sin^2(\theta_k-\Omega_0t)-(\sin\phi+\frac{V_A\cos\phi}{\Omega_0})\cos(\theta_k-\Omega_0t)\bigg).
\end{equation}
Averaging over one driving period gives
\begin{equation}
\langle\Psi_\pm^\dagger\hat L_z\Psi_\pm\rangle_\mathrm{cycle}=\pm\frac{V_+-V_-}{\Omega_0}\sin\phi.
\end{equation}
Using the definitions in Appendix~\ref{sec:effective} gives
\begin{equation}
\langle\Psi_\pm^\dagger(\kk)\hat L_z\Psi_\pm(\kk)\rangle_\mathrm{cycle}=\pm\frac{eA_0}{2\Omega_0}\frac{\d\Delta_0(k)}{\d k}\frac{eA_0\Delta_0(k)\calA_{+-}(k)}{\sqrt{(eA_0\Delta_0(k)\calA_{+-}(k))^2+(\Omega_0-\Delta_0(k))^2}}\sgn(k-k_0).
\end{equation}

The magnetization is
\begin{equation}\label{eq:m-res-raw}
\begin{aligned}
m&=\mu_B\int_{k_{F+}}^{k_{F-}}\frac{k\d k}{2\pi}\int_0^{2\pi}\frac{\d\theta_k}{2\pi}\left(-\frac{2keA_0}{m_2\Omega_0}\right)\frac{keA_0}{\sqrt{(keA_0)^2+(k_0^2-k^2)^2}}\sgn(k-k_0)\\
&=\frac{\mu_B eA_0}{\pi m_2\Omega_0}\int_{k_{F+}}^{k_{F-}} \frac{k^3eA_0}{\sqrt{(keA_0)^2+(k_0^2-k^2)^2}}\sgn(k_0-k)\d k\\
&=\frac{\mu_B e^2A_0^2}{\pi m_2\Omega_0}\cdot\frac12(F(k_0)-F(k_{F+})-F(k_{F-})+F(k_0))\\
&=\frac{\mu_B e^2A_0^2}{2\pi m_2\Omega_0}(2F(k_0)-F(k_{F+})-F(k_{F-})).
\end{aligned}
\end{equation}
Here,
\begin{equation}\label{eq:Fk-res}
\begin{aligned}
F(k)&=\int \frac{2k^3}{\sqrt{(keA_0)^2+(k_0^2-k^2)^2}}\d k\\
&=\sqrt{(k^2-k_0^2)^2+(eA_0k)^2}+\left(k_0^2-\frac{(eA_0)^2}{2}\right)\ln\left(2\sqrt{(k^2-k_0^2)^2+(eA_0k)^2}+2(k^2-k_0^2)+(eA_0)^2\right).
\end{aligned}
\end{equation}
Therefore
\begin{equation}
\begin{aligned}
F(k_0)&=eA_0k_0+k_0^2\ln\left(2eA_0k_0+(eA_0)^2\right)+O(A_0^2),\\
F(k_{F+})&=k_0^2-k_{F+}^2+k_0^2\ln\left(2(k_{F+}^2-k_0^2)+2\left(k_0^2-k_{F+}^2+\frac{(eA_0k_{F+})^2}{2(k_0^2-k_{F+}^2)}\right)+(eA_0)^2\right)+O(A_0^2)\\
&=k_0^2-k_{F+}^2+k_0^2\ln\frac{(eA_0k_0)^2}{k_0^2-k_{F+}^2}+O(A_0^2),\\
F(k_{F-})&=k_{F-}^2-k_0^2+k_0^2\ln(4(k_{F-}^2-k_0^2))+O(A_0^2),
\end{aligned}
\end{equation}
and
\begin{equation}
\begin{aligned}
2F(k_0)-F(k_{F+})-F(k_{F-})&=k_0^2\ln\frac{(2eA_0k_0)^2}{4(k_{F-}^2-k_0^2)\cdot\frac{(eA_0k_0)^2}{k_0^2-k_{F+}^2}}-k_{F-}^2+k_{F+}^2+2eA_0k_0+O(A_0^2)\\
&=k_0^2\ln\frac{k_0^2-k_{F+}^2}{k_{F-}^2-k_0^2}-k_{F-}^2+k_{F+}^2+2eA_0k_0+O(A_0^2).
\end{aligned}
\end{equation}
Substituting into \cref{eq:m-res-raw} gives \cref{eq:m-weak-drive}.

\section{Solution of the Floquet QBE}\label{sec:QBE-solution}
\addtoappendixtoc{Solution of the Floquet QBE}

For the AC-driven system, the Hamiltonian in \cref{eq:QBE_band} is the periodic $H_0^\mathrm{band}(\kk,t)$ in \cref{eq:FloquetH_bandbasis}. We solve this Floquet QBE in the periodic modes
\begin{equation}
|\Phi_\pm(\kk,t)\rangle=e^{iE_\pm(\kk) t}|\Psi_\pm(\kk,t)\rangle.
\label{eq:stat_Floquet_basis}
\end{equation}
They satisfy $(i\partial_t+E_m(\kk))|\Phi_m(\kk,t)\rangle=H_0^\mathrm{band}(\kk,t)|\Phi_m(\kk,t)\rangle$. \cref{eq:Floquet_eigvecs} gives
\begin{equation}\label{eq:V_mat}
|\Phi_m(\kk,t)\rangle=\sum_{s=\pm} V_{sm}(\kk,t) |\psi_s(\kk)\rangle,\qquad \mathbf{V}(\kk,t)=\begin{bmatrix}
\cos(\phi/2)e^{i\theta_k}(1+iK_+-L_1) & -\sin(\phi/2)e^{i\theta_k}(1+iK_++L_2^*)\\
\sin(\phi/2)e^{i\Omega_0t}(1+iK_-+L_2) & \cos(\phi/2)e^{i\Omega_0t}(1+iK_--L_1^*)
\end{bmatrix}.
\end{equation}
Here
\begin{equation}
\begin{gathered}
\phi(k)=\arctan\frac{2keA_0}{m_2\Omega_0-2k^2},\quad\phi\in[-\frac{\pi}{2},\frac{\pi}{2}],\qquad K_\pm=2\frac{(m_1^{-1}\pm m_2^{-1})keA_0}{\Omega_0}\sin(\theta_k-\Omega_0t),\\
L_1=\frac{keA_0}{2m_2\Omega_0}\tan(\phi/2)e^{-2i(\theta_k-\Omega_0t)},\qquad L_2=\frac{keA_0}{2m_2\Omega_0}\cot(\phi/2)e^{2i(\theta_k-\Omega_0t)}.
\end{gathered}
\end{equation}

In this basis, $\rho_{mm'}^\mathrm{F}(\kk,t)=\langle\Phi_m(\kk,t)|\hat\rho|\Phi_{m'}(\kk,t)\rangle$, and the Floquet QBE becomes
\begin{equation}\label{eq:Floquet_QBE}\frac{\d\rho_{mm'}^\mathrm{F}(\kk,t)}{\d t}+i(E_m(\kk)-E_{m'}(\kk))\rho_{mm'}^\mathrm{F}(\kk,t)=e\EE_\mathrm{dc}\cdot(\nabla_\kk\rho_{mm'}^\mathrm{F}(\kk,t)-i[\calA^\mathrm{F}(\kk,t),\rho^\mathrm{F}(\kk,t)]_{mm'}).
\end{equation}
The Floquet Berry connection is
\begin{equation}
\begin{aligned}
\calA^\mathrm{F}_{mm'}(\kk,t)=\sum_{ss'} V_{ms}^\dagger(\kk,t)\calA_{ss'}(\kk)V_{s'm'}(\kk,t)+i\sum_s V_{ms}^\dagger(\kk,t)\nabla_\kk V_{sm'}(\kk,t)=\mathbf{V}^\dagger\calA\mathbf{V}+i\mathbf{V}^\dagger\nabla_\kk\mathbf{V}.
\end{aligned}
\end{equation}
The first term corresponds to a unitary transformation of the static Berry connection, and the second term arises from the $\kk$-dependence of the transformation matrix itself.

Substitution of \cref{eq:V_mat} gives
\begin{equation}
\begin{aligned}
\calA^\mathrm{F}(\kk,t)=\sum_{M=-\infty}^\infty \begin{bmatrix}
\XX_M(k) & \YY_M(k)\\
\YY_{-M}^*(k) & -\XX_M(k)
\end{bmatrix}\cdot e^{iM(\theta_k-\Omega_0t)}.
\end{aligned}
\end{equation}
The components entering the linear response are
\begin{equation}\label{eq:Y_def}
\begin{aligned}
\YY_1&=-i\frac{eA_0\sin\phi}{m_2\Omega_0}\hat{e}_k+\left[-\sin^2\frac{\phi}{2}+\frac{ekA_0\sin\phi}{2m_2\Omega_0}\right]\frac{\hat{e}_\theta}{k}+O(A_0^2),\\
\YY_{-1}&=i\frac{eA_0\sin\phi}{m_2\Omega_0}\hat{e}_k+\left[\cos^2\frac{\phi}{2}+\frac{ekA_0\sin\phi}{2m_2\Omega_0}\right]\frac{\hat{e}_\theta}{k}+O(A_0^2),\\
\YY_0&=-\frac12i\frac{\partial\phi}{\partial k}\hat{e}_k+\frac12\sin\phi\frac{\hat{e}_\theta}{k}+O(A_0),\\
\YY_{M\neq0,\pm1}&=O(A_0).
\end{aligned}
\end{equation}
The density matrix is expanded as
\begin{equation}
\rho^\mathrm{F}(\kk,t)=\sum_{M=-\infty}^\infty \begin{bmatrix}
F_M(\kk)+f_M(\kk) & g_M(\kk)\\
g_{-M}^*(\kk) & F_M(\kk)-f_M(\kk)
\end{bmatrix}\cdot e^{iM(\theta_k-\Omega_0t)}.
\end{equation}
Substituting these two equations into the Floquet QBE \cref{eq:Floquet_QBE} yields a set of coupled algebraic equations for the functions $F_M(\kk)$, $f_M(\kk)$, and $g_M(\kk)$:
\begin{equation}
\begin{aligned}
(-\tau_{F,M}^{-1}-iM\Omega_0)F_M=&eE_\mathrm{dc}\bigg[\cos\theta_k\frac{\partial F_M}{\partial k}-\frac{\sin\theta_k}{k}\bigg(\frac{\partial F_M}{\partial\theta_k}+iMF_M\bigg)\bigg],\\
(-\tau_{f,M}^{-1}-iM\Omega_0)f_M=&eE_\mathrm{dc}\bigg[\cos\theta_k\frac{\partial f_M}{\partial k}-\frac{\sin\theta_k}{k}\left(\frac{\partial f_M}{\partial\theta_k}+iMf_M\right)\bigg]\\
&-i\sum_{M'=-\infty}^\infty e\EE_\mathrm{dc}\cdot(g_{-(M-M')}^*\YY_{M'}-g_{M-M'}\YY_{-M'}^*),\\
i(\Delta-M\Omega_0) g_M=&eE_\mathrm{dc}\bigg[\cos\theta_k\frac{\partial g_M}{\partial k}-\frac{\sin\theta_k}{k}\left(\frac{\partial g_M}{\partial\theta_k}+iMg_M\right)\bigg]\\
&-2i\sum_{M'=-\infty}^\infty e\EE_\mathrm{dc}\cdot(g_{M-M'}\XX_{M'}-f_{M-M'}\YY_{M'}).
\label{components_QBE}
\end{aligned}
\end{equation}
The phenomenological relaxation rates $\tau_{F,M}^{-1}$ and $\tau_{f,M}^{-1}$ regularize the longitudinal responses and describe the scattering required for a steady state under a DC drive. Interband coherence determines the orbital Hall responses, which remain well-defined in the ideal clean limit.

For linear response to $E_\mathrm{dc}$, the zeroth-order density matrix is diagonal in the periodic Floquet-mode basis \cref{eq:stat_Floquet_basis}:
\begin{equation}
F_M^{(0)}=\frac{\delta_{M0}}{2}(n_++n_-),\qquad f_M^{(0)}=\frac{\delta_{M0}}{2}(n_+-n_-),\qquad g_M^{(0)}=0,
\end{equation}
where $n_\pm$ denote the occupations assigned to the two Floquet modes. The frozen zero-temperature occupations are $n_\pm=f(\epsilon_\pm(k))$. Thus $n_--n_+$ is given by \cref{eq:occupation-window}. This approximation neglects drive-induced heating and bath-dependent redistribution. To first order in $E_\mathrm{dc}$, the density-matrix components are
\begin{equation}
\begin{aligned}
F_M&=\frac{\delta_{M0}}{2}(n_++n_-)-\frac{\delta_{M0}}{2}\tau_{F,0} eE_\mathrm{dc}\cos\theta_k \cdot\partial_k(n_++n_-)+O(E_\mathrm{dc}^2),\\
f_M&=\frac{\delta_{M0}}{2}(n_+-n_-)-\frac{\delta_{M0}}{2}\tau_{f,0} eE_\mathrm{dc}\cos\theta_k \cdot\partial_k(n_+-n_-)+O(E_\mathrm{dc}^2),\\
g_M&=\frac{e\EE_\mathrm{dc}\cdot \YY_M}{\Delta-M\Omega_0}(n_+-n_-)+O(E_\mathrm{dc}^2).
\end{aligned}
\end{equation}
Substituting \cref{eq:Y_def} and transforming back to the band basis by
\begin{equation}
\rho^\mathrm{band}(\kk,t)=\mathbf{V}(\kk,t)\rho^\mathrm{F}(\kk,t)\mathbf{V}^\dagger(\kk,t),
\end{equation}
and averaging over one driving period $T=2\pi/\Omega_0$ yields
\begin{equation}\label{offdiag}
\begin{aligned}
&\Re\langle\rho_{+-}^\mathrm{band}(\kk)\rangle_\mathrm{cycle}=\frac{keA_0}{m_2\Omega_0}\sin\phi(n_--n_+)-\cos\theta_keE_\mathrm{dc}\frac{keA_0\sin\phi}{m_2\Omega_0}\tau_{f,0}\partial_k(n_--n_+)\\
+&\sin\theta_keE_\mathrm{dc}\frac{(n_--n_+)}{k}\bigg(\frac{\Omega_0\cos\phi-\Delta(1+\cos^2\phi)/2}{\Omega_0^2-\Delta^2}-\frac{keA_0\sin\phi}{m_2\Omega_0}\frac{\cos\phi}{\Delta}\bigg).
\end{aligned}
\end{equation}

\section{Evaluation of the conductivities}\label{sec:integral}
\addtoappendixtoc{Evaluation of the conductivities}

For the longitudinal orbital conductivity, the derivative of the occupation difference reduces the radial integral to the two Fermi contours:
\begin{equation}
\partial_k(n_+-n_-)=\delta(k-k_{F-})-\delta(k-k_{F+}),
\end{equation}
which gives
\begin{equation}
\begin{aligned}
\sigma_{L_z}^L=\frac{J_{L_z}^L}{E_\mathrm{dc}}&=\frac{e^2\tau_{f,0}A_0}{\pi m_1m_2\Omega_0}(k_{F-}^3\sin\phi(k_{F-})-k_{F+}^3\sin\phi(k_{F+}))\\
&=-\frac{e^3\tau_{f,0}A_0^2}{\pi m_1m_2\Omega_0}\left(\frac{k_{F-}^4}{\sqrt{(k_{F-}eA_0)^2+(k_0^2-k_{F-}^2)^2}}+\frac{k_{F+}^4}{\sqrt{(k_{F+}eA_0)^2+(k_0^2-k_{F+}^2)^2}}\right)\\
&=-\frac{e^3\tau_{f,0}A_0^2}{\pi m_1m_2\Omega_0}(G_L(k_{F-})+G_L(k_{F+})),\qquad G_L(k)=\frac{k^4}{\sqrt{(k_0^2-k^2)^2+(eA_0k)^2}}.
\end{aligned}
\end{equation}

For the orbital Hall conductivity, the two terms in the last two lines of \cref{eq:offdiag} can be simplified as
\begin{equation}
\frac{\Omega_0\cos\phi-\Delta(1+\cos^2\phi)/2}{\Omega_0^2-\Delta^2}=\frac{m_2}{2}\frac{k_0^2\cdot\frac{k_0^2-k^2}{\sqrt{(k_0^2-k^2)^2+(keA_0)^2}}+\frac{\sqrt{(k_0^2-k^2)^2+(keA_0)^2}}{2}\left(1+\frac{(k_0^2-k^2)^2}{(k_0^2-k^2)^2+(keA_0)^2}\right)}{k_0^4-(k_0^2-k^2)^2-(keA_0)^2}\sgn(k_0-k),
\end{equation}
and
\begin{equation}
\frac{\sin\phi\cos\phi}{\Delta}=\frac{m_2}{2}\frac{keA_0\cdot(k_0^2-k^2)}{((keA_0)^2+(k_0^2-k^2)^2)^{3/2}}\sgn(k_0-k),
\end{equation}
Combining these identities and splitting the radial integral at $k_0$, as required by $k_{F+}<k_0<k_{F-}$, gives
\begin{equation}\label{eq:SigmaH-res}
\sigma_{L_z}^H=\frac{J_{L_z}^H}{E_\mathrm{dc}}=\frac{m_2e}{2\pi m_1}\left[\int_{k_{F+}}^{k_0}\frac{\d G_H}{\d k}\d k-\int_{k_0}^{k_{F-}}\frac{\d G_H}{\d k}\d k\right]=\frac{m_2e}{2\pi m_1}[2G_H(k_0)-G_H(k_{F-})-G_H(k_{F+})].
\end{equation}
where
\begin{equation}\label{eq:GHk-int}
\begin{aligned}
G_H(k)&=\int \bigg[\frac{k_0^2(k_0^2-k^2)+(k_0^2-k^2)^2+\frac12(keA_0)^2}{\sqrt{(k_0^2-k^2)^2+(keA_0)^2}(k_0^4-(k_0^2-k^2)^2-(keA_0)^2)}-\frac{(eA_0)^2}{2k_0^2}\frac{k^2(k_0^2-k^2)}{((k_0^2-k^2)^2+(keA_0)^2)^{3/2}}\bigg]\cdot k\d k.
\end{aligned}
\end{equation}
Explicit integration gives
\begin{equation}\label{eq:GHk-finalres}
\begin{aligned}
G_H(k)=&\frac{2 k_0^2}{2k_0^2-(eA_0)^2}\tanh^{-1}\frac{k^2-\sqrt{(k^2-k_0^2)^2+(eA_0k)^2}}{k_0^2}\\
+&\frac{(eA_0)^4}{2k_0^2(2k_0^2-(eA_0)^2)}\tanh^{-1}\frac{2k_0^2-k^2-(eA_0)^2+\sqrt{(k^2-k_0^2)^2+(eA_0k)^2}}{k_0^2}\\
-&\frac{2k_0^2-(eA_0)^2}{4k_0^2}\ln\left[2\sqrt{(k^2-k_0^2)^2+(eA_0k)^2}+2(k^2-k_0^2)+(eA_0)^2\right]\\
+&\frac{(eA_0)^2}{2k_0^2}\frac{k_0^4-3(k k_0)^2+(keA_0)^2}{(4k_0^2-(eA_0)^2)\sqrt{(k^2-k_0^2)^2+(eA_0k)^2}}.
\end{aligned}
\end{equation}
Therefore
\begin{equation}
\begin{aligned}
G_H(k_0)&=\frac{2k_0^2}{2k_0^2-(eA_0)^2}\tanh^{-1}\left(1-\frac{eA_0}{k_0}\right)-\frac{2k_0^2-(eA_0)^2}{4k_0^2}\ln(2eA_0k_0+(eA_0)^2)+\frac{eA_0((eA_0)^2-2k_0^2)}{2k_0(4k_0^2-(eA_0)^2)}+O(A_0^4)\\
&=-\ln(eA_0)-\frac{3eA_0}{4k_0}+O(A_0^2),\\
G_H(k_{F+})&=\frac12\ln\frac{k_{F+}^2}{k_0^2-k_{F+}^2}-\frac12\ln\frac{(eA_0)^2k_0^2}{k_0^2-k_{F+}^2}+O(A_0^2)\\
&=\ln\frac{k_{F+}}{eA_0k_0}+O(A_0^2),\\
G_H(k_{F-})&=\frac12\ln\frac{4k_0^2(k_{F-}^2-k_0^2)}{(eA_0)^2k_{F-}^2}-\frac12\ln\left[4(k_{F-}^2-k_0^2)\right]+O(A_0^2)\\
&=\ln\frac{k_0}{eA_0k_{F-}}+O(A_0^2),
\end{aligned}
\end{equation}
and
\begin{equation}
2G_H(k_0)-G_H(k_{F+})-G_H(k_{F-})=\ln\frac{k_{F-}}{k_{F+}}-\frac{3eA_0}{2k_0}+O(A_0^2).
\end{equation}
Substituting into \cref{eq:SigmaH-res} gives \cref{eq:OHC-weak-drive}.

\end{document}